\documentclass{aa}
\usepackage{graphics}
\usepackage{epsfig}

\def\msol{M_\odot}
\def\mmin{m_{\rm min}}
\def\mb{M_{\rm B}}
\def\mv{M_{\rm V}}
\def\mi{M_{\rm I}}
\def\mj{M_{\rm J}}
\def\mh{M_{\rm H}}
\def\mk{M_{\rm K}}
\def\wi{W_{\rm I}}
\def\te{T_{\rm eff}}
\def\lsol{L_\odot}
\def\simgr{\,\hbox{\hbox{$ > $}\kern -0.8em \lower 1.0ex\hbox{$\sim$}}\,}
\def\simle{\,\hbox{\hbox{$ < $}\kern -0.8em \lower 1.0ex\hbox{$\sim$}}\,}
\def\beq{\begin{equation}}
\def\eeq{\end{equation}}

\def\aj{AJ}                  
\def\apj{ApJ}                 
\def\apjs{ApJS}               
\def\aap{A\&A}                
\def\aaps{A\&AS}             
\def\mnras{MNRAS}             

\begin{document}

\thesaurus{08.05.03, 08.22.1}

\title{Period - magnitude relationships in BVIJHK-Bands
for fundamental mode and first overtone Cepheids}

\author{Isabelle Baraffe\inst{1,2} \and Yann Alibert\inst{1}}

\offprints{I. Baraffe}

\institute{C.R.A.L (UMR 5574 CNRS), 
 Ecole Normale Sup\'erieure, 69364 Lyon
Cedex 07, France
\and Max-Planck Institut f\"ur Astrophysik, Karl-Schwarzschildstr.1,
D-85748 Garching, Germany
\\ email: ibaraffe, yalibert@ens-lyon.fr
}

\date{Received /Accepted}

\titlerunning{Period - magnitude relationships}
\authorrunning{I. Baraffe and Y. Alibert} 
\maketitle

\begin{abstract}

We present theoretical period - magnitude relationships for Cepheids
in different filters for fundamental and first overtone
pulsators, completing the work by Alibert et al. (1999). 
The results are provided for different metallicities 
characteristic of the Magellanic Clouds
and the Milky Way. In contrast
to the fundamental mode, we find a small metallicity effect
on the period - luminosity relationship for the first overtone,
due to the sensitivity of the period ratio $P_1$/$P_0$
with metallicity.
Comparison is made with observations from OGLE and EROS
in the Small and Large Magellanic Clouds. We emphasize the
constraint on theoretical predictions 
provided by the combination of both  fundamental
and first overtone observed sequences. 
We obtain 
excellent agreement between models and data in a $\log P$ - $\wi$
(Wesenheit index)
diagram for a distance modulus for the LMC $\mu_0$ = 18.60 - 18.70.
We analyse the uncertainties of the fundamental
period - magnitude relationships and the consequences
on distance determination. We show that an arbitrary shift
of the instability strip by 350 K in $\te$ yields up
to 0.45 mag effect on $\mv$ at a given period, whereas
the effect is less than 0.1 mag in the $K$-band. Using
recent near-IR observations in the Large Magellanic Cloud
and our $P$ - $\mk$ relationship,
we derive a distance modulus for the LMC in agreement with the value
based on $\wi$ data. 

\end{abstract}

\section{Introduction}

Over the last five years, a wealth of data for Cepheids was
collected by
several observational projects (HST key project, HIPPARCOS,
microlensing experiments EROS, MACHO, OGLE, etc...). 
This fantastic amount of data allows unprecedented
opportunities to understand the properties of these variable stars.
One of the relevance of  Cepheids, and the main goal
of the HST key project (cf. Freedman 2000, and references therein),
is the determination of local
extragalactic distance scales and consequently of the Hubble
constant $H_0$. 
The period - luminosity relationship (PL) of 
 Cepheids is the conerstone of such purpose.
An important contribution to the collection of data is provided by microlensing
search projects toward the Magellanic Clouds (EROS: Renault et al. 1996; OGLE: 
Udalski, Kubiak and Szymanski 1997; MACHO: Alcock et al. 1997; and references therein).
 The recent release to the public domain of their
Cepheid Catalog (EROS: Afonso et al. 2000; OGLE: Udalski
et al. 1999a,b)
provides large, homogeneous samples of high quality data which in principle
allow a deep analysis of the period-luminosity relationships. In particular,
a key question related to the PL relationship is its
dependence on metallicity. This is of fundamental importance 
for extragalactic distance scales, usually based on
 an universal PL relationship. 
A controversy still exists regarding metallicity effects, both
on the observational and theoretical viewpoints (see Alibert et al. 1999 and
references therein). Another notorious debate concerns the distance
to the LMC, which is  an important source of uncertainty in the
determination
of $H_0$ (see Freedman 2000).
Although PL relationships  used for distance estimate are based on 
fundamental (F)  pulsators, the afore-mentioned catalogs provide also 
statistically significant samples of 
first overtone (1H) Cepheids. Because of the distinction of
F and 1H pulsator sequences in period - magnitude diagrams,
1H provides additional constraints on theoretical models and
on distance determinations based on PL relationships. 

In a recent paper, Alibert et al. (1999) performed self-consistent
stellar evolution and linear stability analysis calculations for Cepheids.
Their analysis was essentially devoted to fundamental mode pulsators
and their period-magnitude-color relationships
as a function of metallicity. The analysis of their results for first
overtone pulsators and the comparison with observations
were hampered at that time by the lack of significant samples of
data. 
The aim of the present paper is to complete the work of Alibert et al. (1999)
for first overtone pulsators and test their results
against the observational constraints provided by {\it both}
F and 1H pulsators. In section 2, we derive PL relationships
for 1H and in \S 3 we
compare the results to observations. 
In \S 4, we briefly discuss 
the uncertainties of our
period - magnitude relationships and consequences on distance
determinations.

\section {Period - magnitude relationships for first overtone pulsators}

The description of the stellar evolution and pulsation calculations are
given in details in Alibert et al. (1999). We recall the main ingredients:
(i) evolutionary models for Cepheids are constructed with the Lyon
evolutionary code from 3 to 12 $\msol$. We consider various
initial compositions
$(Z, Y) = (0.02, 0.28)$, $(0.01, 0.25)$
and  $(0.004, 0.25)$ \footnote {$Z$ is the metal mass fraction
 and $Y$ the helium mass fraction}, representative of respectively the
Galactic, the Large Magellanic (LMC)  and the Small Magellanic (SMC) clouds environments.  Evolutionary calculations do not include core overshooting.
(ii) A linear non-adiabatic stability analysis is performed on the complete evolutionary
models along the evolutionary tracks. This provides consistent mass-age-period-luminosity
relations. 
(iii) Static atmosphere models and their corresponding synthetic spectra
are calculated for the same compositions used in (i), providing magnitudes and colors
for a given ($Z$, $M$, $L$, $\te$). 

\subsection{Period - luminosity and Period - magnitude relationships in
different filters.}

In order to  derive statistical PL relationships, a mean position 
is assigned to each mass, according to the time spent in different 
locations in the instability strip (IS) and a linear
least-square fit to these points is derived. As shown in 
Alibert et al. (1999), the mean position of a given stellar mass,
accounting for  its evolutionary time, 
is roughly located in the middle of the IS. 

The minimum masses $\mmin$  undergoing a blue loop 
in the 1H instability strip are 3.25 $\msol$, 4 $\msol$ and
5 $\msol$ for respectively $Z$ = 0.004, 0.01 and 0.02. In order to
avoid biases due to the change of slope predicted near these minimum
masses
 (see Alibert et al. 1999, their \S 3.4), we exclude in the present analysis
$\mmin$ to
derive the mean PL relationships. Because of the reduction of the blue
loop extension toward lower masses,  
the slope of the PL relationships for 1H can be affected if the fit
is derived down to $\mmin$, yielding steeper relations.
Bauer et al. (1999) observed such a change of slope for fundamental
pulsators in the SMC, but did not observe it for 1H, although
predicted by the models (see the
discussion
by Alibert et al. 1999,  \S 4.2.3).
The relationships given in Tables \ref{tab1}
and \ref{tab2} 
are then derived from 3.5 $\msol$ for $Z$ = 0.004, 4.25 $\msol$ for $Z$ = 0.01
and 5.5 $\msol$ for $Z$ = 0.02. 
Note that
Alibert et al. (1999) derived preliminary PL relationships for 1H based
on the whole range of masses unstable in the IS (including $\mmin$)
and obtained slightly different
results than  in Table \ref{tab1}. 

\begin{table}
\caption{Coefficients of the $\log \, P$ (in days) - $\log \, L/L_{\odot}$  relationships
 (slope, zero-point)
for  first overtone pulsators
as a function of metallicity.}
\begin{tabular}{lc}
\hline\noalign{\smallskip}
$Z$=0.02  & (1.247, 2.659))  \\
$Z$=0.01  & (1.252, 2.646)  \\
$Z$=0.004 & (1.257, 2.613) \\
\hline
\end{tabular}
\label{tab1}
\end{table}

The period - magnitude
relationships in different filters are given in Table \ref{tab2} for
F and 1H modes. 
$BVRI$ magnitudes are based on the Johnson-Cousins system 
(Bessell 1990)
and  $JHK$ magnitudes are defined in the CIT system (Leggett
1992). Table \ref{tab2} displays also the reddening-free Wesenheit
index $W_{\rm I} \, = \, I \, - \, 1.55(V-I)$ (Madore and Freedman 1991),
 with the coefficient 1.55 ($= \, A_{\rm I}/E(V-I)$) resulting from
standard interstellar extinction curves ({\it e.g.,} Schlegel et al. 1998).

\begin{table*}
\caption{Coefficients of the $\log \, P$ (in days) - Magnitude relationships
 (slope, zero-point)
for fundamental mode (upper row) and first overtone (lower row)
as a function of metallicity.}
\begin{tabular}{lccccccc}
\hline\noalign{\smallskip}
$Z$ & $M_B$  &$M_V$  & $M_I$ & $M_J$ & $M_H$  & $M_K$ & $W_I$    \\
\noalign{\smallskip}
\hline\noalign{\smallskip}
0.02 & (-2.658,-0.648)&
(-2.905,-1.183) & (-3.102,-1.805) & (-3.256,-2.183) &
 (-3.346,-2.428) & (-3.367,-2.445) & (-3.407,-2.770) \\
 & (-2.669,-1.561) & (-2.948,-1.990) & (-3.171,-2.526) & (-3.333,-2.844) &
(-3.434,-3.042)& (-3.453,-3.057)& (-3.515,-3.355) \\
\hline
0.01  & (-2.712,-0.655)
& (-2.951,-1.153) & (-3.140,-1.769) & (-3.286,-2.157) &
 (-3.377,-2.411) & (-3.395,-2.428) & (-3.433,-2.725) \\
 & (-2.699,-1.515)& (-2.972,-1.923) &(-3.192,-2.463)& (-3.356,-2.795)&
(-3.461,-3.006)& (-3.481,-3.019)&(-3.534,-3.300) \\
\hline
0.004 & (-2.707,-0.633) 
& (-2.939,-1.081) & (-3.124,-1.686) & (-3.262,-2.076) &
 (-3.351,-2.336) & (-3.369,-2.350) & (-3.411,-2.623) \\
 & (-2.765,-1.430)&(-3.01,-1.807) & (-3.212,-2.341)& (-3.376,-2.678)
& (-3.478,-2.898) & (-3.501,-2.906) & (-3.545,-3.170)\\
\hline
\end{tabular}
\label{tab2}
\end{table*}

\subsection{Effects of metallicity}

\begin{figure}
\psfig{file=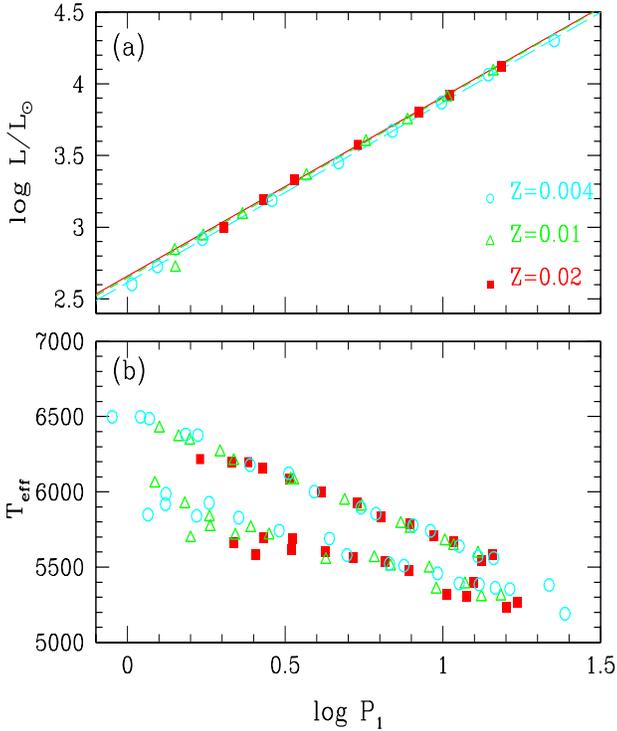,height=110mm,width=88mm}
\caption{{\bf (a)} Log $P$ (in days) - $L$ diagram for 1H pulsators and
 different metallicities. The symbols
indicate the mean position of each mass in the IS as a function of $Z$.
The mean $PL$ relationships are plotted for $Z$ =0.02 (solid line), $Z$=0.01
(dashed line) and $Z$=0.004 (long dashed-line). 
{\bf (b)} Location of the blue and red edges of the 1H IS in a 
$\log \, P$ - $\te$ diagram. The symbols correspond to the same
$Z$ as in (a). The minimum masses undergoing a blue loop in the IS
are 3.25 $\msol$, 4 $\msol$ and
5 $\msol$ for respectively $Z$ = 0.004, 0.01 and 0.02.
}
\label{fig1}
\end{figure}

Figure \ref{fig1} displays the mean $\log P_1$ - $\log L/\lsol$
 relationships (Fig. \ref{fig1}a)
 and the blue and red edges in a $\log P_1$ - $\te$ diagram
  (Fig. \ref{fig1}b) as a function of $Z$.
Excluding the minimum mass undergoing a blue loop in the IS, which
decreases with $Z$, there is no noticeable
effect of $Z$ on the location and width of the IS in Fig. 1b. 
The periods corresponding to $\mmin$ in Fig. \ref{fig1}
are $\log \, P_1 \sim$ 0.3, 0.1 and 0 for respectively
$Z$ = 0.02, 0.01 and 0.004.
We note however
a small effect of $Z$ on the PL relationship, with $\log \, L$
increasing by 0.04 for $Z$ increasing from 0.004 to 0.02
at a given $P_1$.

In the following, our analysis  is restricted to
 period   $ \log \, P_1 \, \ge$ 0.3, since $\log \, P_1$ = 0.3
is the minimum period for $Z$=0.02 1H pulsators undergoing a blue loop
(see  Fig. \ref{fig1}).
An inspection of the relationships given in
Table \ref{tab1} shows that the slopes  hardly depend on $Z$, but
the zero points are  affected by
 metallicity. Although small, the increase of $L$ with $Z$
translates in brighter magnitudes at a given $P_1$, with the largest
effects in the $VIJ$ bands and the maximum effect in the $I$ band.
We note that $\mv$, $\mi$ and $\mj$ are
brighter by 0.1 - 0.15 mag, $\mh$ and $\mk$ are brighter
by 0.8 - 0.13 mag at a given $P_1$ when $Z$ increases
from 0.004 to 0.02. Between $Z$=0.004 and $Z$ = 0.01 the effect
is $\sim$ 0.1 mag in the $VIJHK$ bands. Interestingly enough, the
effect is the smallest in the $B$ band, with a difference of $\mb$
less than 0.1 mag between $Z$ = 0.004 and $Z$ = 0.02. Opposite
trends are found for fundamental mode PL relationships which show
the largest $Z$ effect in the $B$ band, but small
effects (less than 0.1 mag) at longer wavelengths (see Alibert et
 al. 1999,
\S 5). 

This effect of $Z$ on the PL relationship for 1H is related to the behavior
of the period ratio $P_1$/$P_0$ as a function of $Z$. 
This ratio
decreases as $Z$ increases (cf. Baraffe et al. 1998): we find
for $Z$ = 0.004, $P_1$/$P_0$ $\sim$ 0.73 and for
$Z$ = 0.02, $P_1$/$P_0$ $\sim$ 0.7.
Since  for a given $L$,
 $P_0$ is essentially
independent of $Z$, $P_1$ decreases
  because of decreasing
$P_1$/$P_0$ ratio for increasing $Z$. Conversely, for a given $P_1$,
$L$ increases with $Z$.   

 The behavior of $P_1$/$P_0$ with $Z$ can be understood by analysing
the regions in the star which contribute to the period
of each mode. Such regions can be determined by the Epstein weight
functions which are obtained by expressing the eigen frequency of
each mode by an integral over the whole star (Epstein 1950; Cox 1980).
As already noted by Epstein (1950), the greatest contribution to the fundamental
mode period is located between $r/R \sim$ 0.7 and $r/R \sim$ 0.9. For the first overtone, two regions
are important: a first inner region located between $r/R \sim$ 0.4 and 
$r/R \sim$ 0.7 and an outer zone with $r/R \, > 0.85$. These regions are
characteristic of the mass range of interest (5-10 $\msol$). The inner region for 1H includes the opacity peak due to metals at $T \, \sim 2 \, 10^5$ K
and the outer region covers the H-He ionization zone. In the first region
covering the metal opacity peak, an increase of the opacity $\kappa$, related to an increase of
the metallicity, yields an increase of $T$ or
a decrease of the density $\rho$ at fixed $r$ (or $P$). Note that this region is dominantly radiative
and the temperature gradient depends directly on $Z$ since the radiative
gradient is
proportional to $\kappa$. 
Consequently, in this region, the sound speed $C_{\rm s} \propto
(P/\rho)^{1/2} \propto T^{1/2}$ is more sensitive
to the metallicity than in other outer regions and increases with $Z$. The integral of the
sound travel time $\tau_{\rm s} = dr/C_{\rm s}$ through this zone is
thus more sensitive to the metallicity, decreasing as $Z$ increases,
than in the other regions located at 
$r/R > 0.7$.  
The period ratio $P_1$/$P_0$ is related to the ratio
of $\tau_{\rm s}$ through the regions relevant for 1H
to  $\tau_{\rm s}$ in the zones contributing to F. Because of the larger sensitivity to $Z$ of $\tau_{\rm s}$
for 1H, this ratio decreases as $Z$ increases. We checked
our arguments by calculating 
the  integral of $\tau_{\rm s}$ through the different regions of
interest for different cases of $M$, $L$ and $\te$ covering
the instability strip and for different values of $Z$. 



The behavior of $L$ as a function of $P_1$ with $Z$
explains the trends previously mentioned of the fluxes in
different filters.
In particular, it explains the low sensitivity of the $\log \, P_1$ - $\mb$ relationship
to $Z$ because of  the compensating effects of (i) higher $L$ with increasing $Z$ at a given
$P_1$ and (ii) lower $B$-flux for a given $L$ and $\te$ as $Z$ increases,
because of more metallic line absorption in the $B$ bandpass 
(see \S5 of Alibert et al. 1999). 

Finally, because of the maximum effect in the $I$-bandpass, we note
a non negligible effect of $Z$ on the index $\wi$, which differs
by 0.14 mag at long $P_1$ up to 0.18 mag at shorter $P_1$ between
Z=0.004 and 0.02. When $Z$ increases from 0.004 to 0.01, $\wi$
gets brighter by $\sim$ 0.13 mag.

\section{Period - magnitude diagrams: comparison with observations}
 
Comparison with observations are made with the EROS2 and OGLE2 data
for the MC. EROS2 (Afonso et al. 2000) reports 239 1H Cepheids
in the SMC and 113 in the LMC. OGLE2 provides $\sim$ 800 1H pulsators
in the SMC and $\sim$ 500 in the LMC. 

\subsection{SMC Cepheids: $VI$-Bands}

\begin{figure}
\psfig{file=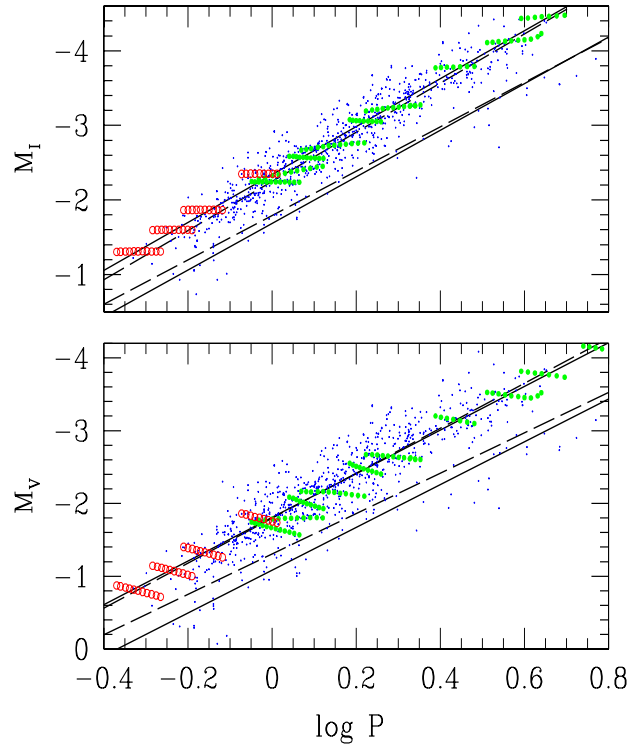,height=110mm,width=88mm}
\caption{Period (in days) - Magnitude diagrams for 1H pulsators in the $VI$ bands for
 models with $Z$=0.004 and SMC observed Cepheids.
 The filled circles correspond to 
unstable 1H modes during core He burning phase from $\mmin$ = 3.25 $\msol$
to 6 $\msol$.
The large open circles correspond
to first crossing  unstable 1H modes for
 3, 3.25, 3.5 and 4 $\msol$. Observations (dots) 
are from OGLE2 (Udalski et al. 1999b).
 The distance modulus is 18.94.
The 2 solid curves are the theoretical mean PL relationships for respectively
F (lower curve) and 1H (upper curve) pulsators. The long-dashed curves are the observed
PL relationship derived by Udalski et al. (1999c) and Udalski (2000)
 for F and 1H pulsators.
}
\label{fig2}
\end{figure}

Figure \ref{fig2} displays the comparison between models with metallicity
$Z$ = 0.004 and the OGLE2 data. EROS2 data are not shown for the sake
of clarity, but are comparable to the OGLE2 data.
We adopt the dereddened
data provided by the OGLE2 catalog, with different reddening corrections
for each fields (see Udalski et al. 1999b for details)
and the same distance modulus $\mu_0 = 18.94$ as used in Alibert et
al. (1999). This  value, suggested 
by Laney and Stobie (1994), yields general
agreement between models  and various
observations for fundamental pulsators (cf. Alibert et al. 1999).
Unstable models during the first crossing phase (large open circles)
and the core He burning
blue loop (filled circles) are indicated (see Alibert et al. 1999 for details).
As shown in Fig. \ref{fig2}, a general good agreement is  found for 1H
pulsators. The  fundamental mode sequence is indicated
by the mean PL relationship,
and compared to the relationship of Udalski et al. (1999c), recently revised by
Udalski (2000). Note
that these authors truncate their SMC sample at $\log P \, < \, 0.4$,
to avoid biases at short periods.
The agreement between
predicted and observed F relationships is thus excellent
for $\log P \, > \, 0.4$. As preliminary noted by Alibert et al. (1999)
and confirmed by the  data,
observations for 1H do not extend above $\log P \, \sim \, 0.6$, which
corresponds to $m \, \sim \, 5 \msol$ with standard 
(no overshooting)
evolutionary models. Note that models including overshooting
would yield a lower mass for such period. Our linear stability analysis
however finds unstable 1H modes (as well as F modes) 
for masses $m > 5 \msol$. Only non-linear
calculations can determine the dominant mode of pulsation. In the same vein, 
models predict that 1H pulsators with $\log P \, \simle \, 0$ are in the first
crossing phase. 
The significant number of such low period pulsators provides
another strong constraint on non-linear calculations which could test the 
following
scenarios: (1) models on the first crossing, characterised by a lower luminosity
than during the He core burning blue loop for a given mass, 
should favor 1H as the dominant mode, and (2) models with $m \, \simgr \, 5 \msol$
on the blue loop should oscillate predominantly in the fundamental mode.

 We note however that in case (1), one may expect a change in the number density
of objects with $\log P \, \simle \, 0$, given the much faster evolutionary
timescale on the first crossing compared to the blue loop
phase. Such a change is not displayed by current observations.
A detailed statistical analysis taking into account evolutionary
timescales, mass functions and eventual observational biases
is required to investigate this point.

\subsection{LMC Cepheids: $VI$-Bands}

\begin{figure}
\psfig{file=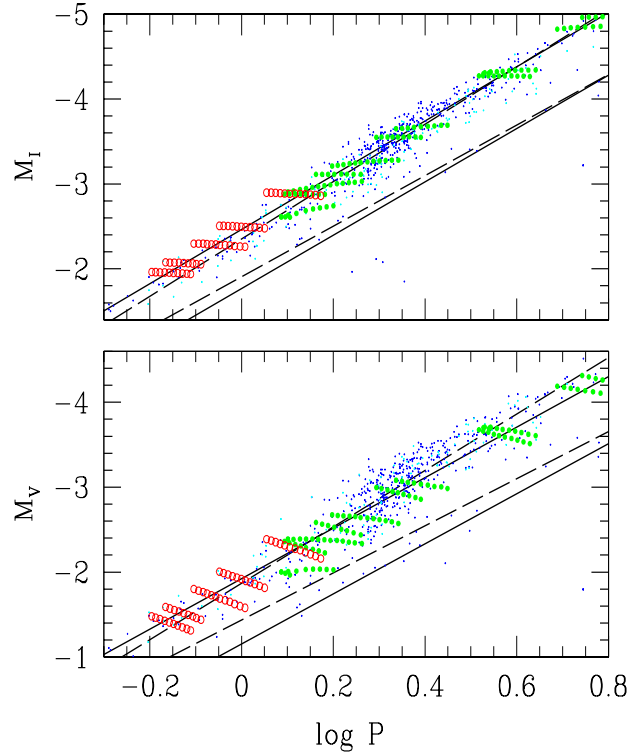,height=110mm,width=88mm}
\caption{Period (in days)  - Magnitude diagrams for 1H pulsators in the $VI$ bands for
 models with $Z$=0.01 and LMC observed Cepheids.
 The filled circles correspond to 
unstable 1H modes during core He burning phase from $\mmin$ = 4 $\msol$
to 7 $\msol$.
The open circles correspond
to first crossing  unstable 1H modes for
 3.875, 4, 4.25, 4.5, 5 $\msol$. Observations (dots) are from OGLE2  
(Udalski et al. 1999a) and EROS2 (Afonso et al. 2000).
 The distance modulus is 18.50.
The 2 solid curves are the theoretical mean PL relationships for respectively
F (lower curve) and 1H (upper curve). The 2 long-dashed curves are the observed
PL relationships derived by Udalski et al. (1999c) and Udalski (2000).
}
\label{fig3}
\end{figure}

Comparison between models with Z=0.01 and observations in the LMC is
shown in Fig. \ref{fig3}. Data are from EROS2 (Afonso et al. 2000) 
and OGLE2 (Udalski et al. 1999a). We adopt the same
reddening corrections as in Udalski et al. (1999a) for the OGLE2 sample, 
which varies from field to field. For the EROS2 data, we adopt the same
reddening correction $E(B-V)$=0.10 as used in Alibert et al. (1999).
We note that adopting a constant $E(B-V)$=0.10 for the OGLE2 data has
no noticeable effect when comparing models and data in Fig. \ref{fig3}.
We adopt the same distance modulus 18.50 as in Alibert et al. (1999).
The theoretical PL relations for F 
and 1H modes are compared to
the observed relations derived by Udalski et al. (1999c) on the LMC sample
limited to  $\log P \, > \, 0.4$, and recently revised by Udalsky (2000). 
The same agreement as for the SMC data is found
and the models reproduce  the width and location of the instability
strip satisfactorily. 
We however note discrepancies between models and observations
for the {\it mean} $P$ - $\mv$ 
relationship for both F and 1H modes. This discrepancy appears
also, but to a lesser extent, for the SMC (cf. Fig. \ref{fig2}) and
is better illustrated in $P$ - $\wi$ diagrams (see next section).

As illustrated in Fig. 3, no 1H pulsators are observed above $\log P \, \sim 0.8$,
corresponding to $m \, \simgr 7 \msol$. 
Note that the lack of 1H pulsators at long periods could also be
interpreted in terms of very small amplitudes below the level of
detection of OGLE or EROS. However, 
there is no obvious decrease of the amplitudes of 1H pulsators
in the LMC, as well as in the SMC (see Afonso et al. 1999),
as the period increases which could support this interpretation.
Below $\log P \, \sim \, 0.1$, 
observations can be explained by first crossing models. 
We therefore find the same type of constraints as derived in
the previous section from
the SMC data on non-linear
calculations for the determination of the dominant pulsating mode.
Such comparisons suggest that the afore-mentioned properties in scenarios
(1) and (2) (see \S 3.1) are intrinsic to 1H pulsators for $Z$ varying from 0.004
to 0.01.
More observations at higher metallicities are currently
required to determine  if these properties
apply also to Galactic Cepheids.   

\subsection{$P$ - $W_{\rm I}$ diagram}

\begin{figure*}
\psfig{file=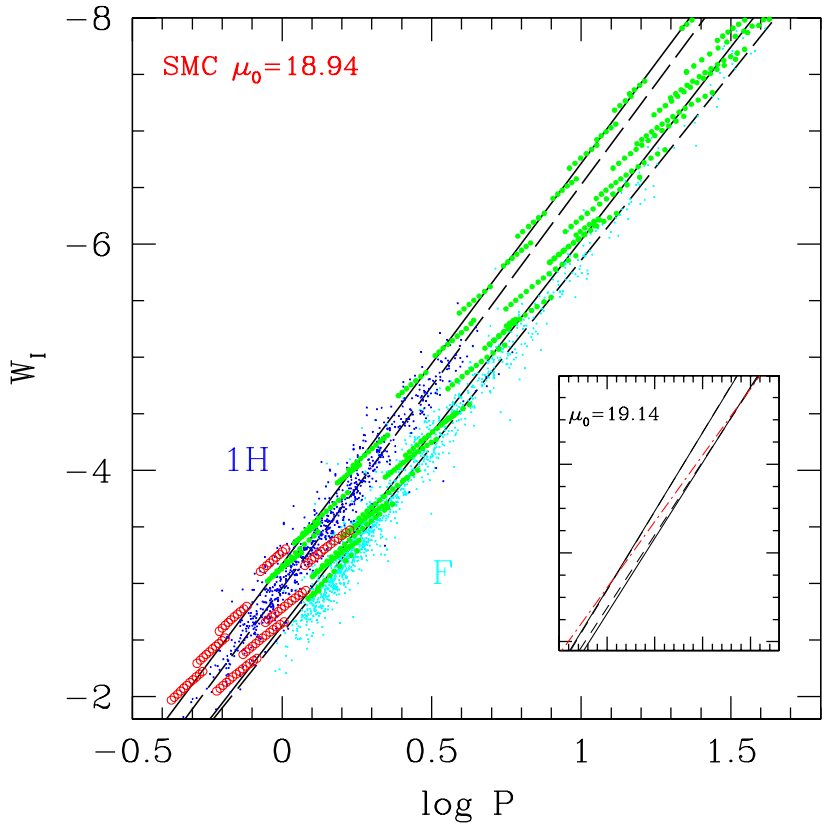,height=120mm,width=120mm}
\caption{$\log \, P$ (in days)  - $\wi$ diagram for F and 1H pulsators  for
 models with $Z$=0.004 and SMC observed Cepheids.
Symbols and curves are the same as in Fig. \ref{fig2}. The inset
displays only the mean relationships with the same
linestyle as in the main figure for a distance modulus $\mu_0$ =
 19.14.
 The dash-dotted
curve corresponds to the F relationship of Caputo et al. (2000).
Note that for 1H, the theoretical and observed mean relationships
are indistinguishable in the inset.
}
\label{fig4}
\end{figure*}

\begin{figure*}
\psfig{file=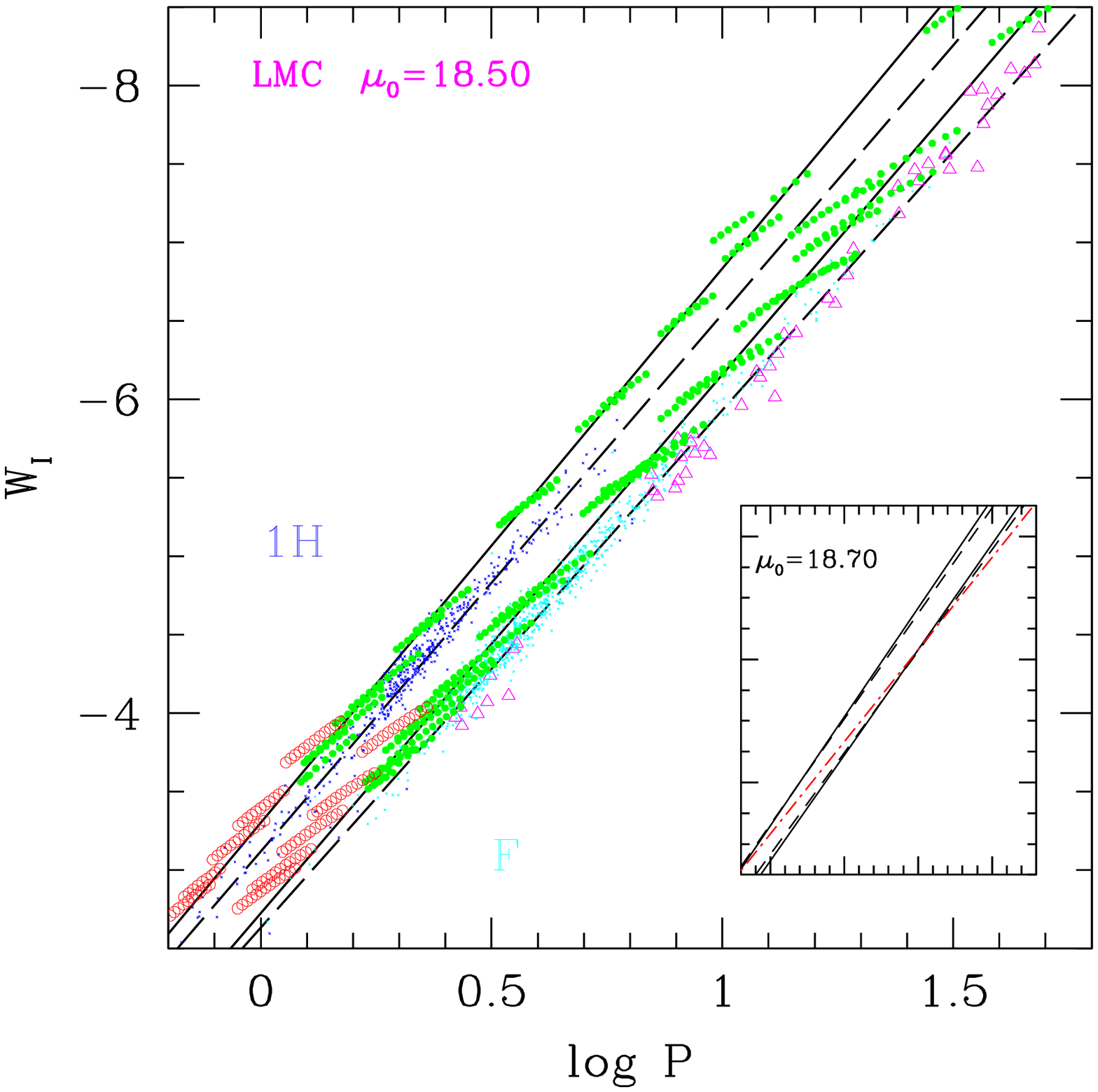,height=120mm,width=120mm}
\caption{$\log \, P$ (in days)  - $\wi$ diagram for F and 1H pulsators  for
 models with $Z$=0.01 and LMC observed Cepheids.
Symbols and curves are the same as in Fig. \ref{fig3}.
The triangles are observations from Gieren et al. (1998).
The inset
displays only the mean relationships with the same
linestyle as in the main figure for a distance modulus $\mu_0$ =
 18.70.
 The dash-dotted
curve corresponds to the F relationship of Caputo et al. (2000).
}
\label{fig5}
\end{figure*}

Figures \ref{fig4} and \ref{fig5} display observations and models for both
F and 1H in a $\log \, P$ - $W_{\rm I}$ diagram, for respectively the SMC and
the LMC. The Wesenheit index $W_{\rm I}$
is an useful quantity  since it is rather insensitive
to the reddening
 and can in principle remove part of the scatter due to differential reddening.
For the LMC, the $Z$ = 0.01 models are also compared 
to observations from Gieren et al. (1998) for F Cepheids.
 Note that Alibert et al. (1999) found
good agreement in different optical and near-infrared $P$ - magnitude
diagrams with the latter sample of data.
As expected, the scatter of data in such diagram is smaller than in the
$V$ and $I$ bands. A first inspection of Figs. \ref{fig4} and  \ref{fig5} 
show a general agreement between models and observations 
 for both SMC and LMC:  the models corresponding to different masses
cover the observed location of both F and 1H Cepheids.
We note that the models are in better agreement with the observed
width of the IS in the LMC, whereas the data in the SMC
show a  larger scatter. This is consistent with the recent analysis
by Groenewegen (2000) who interprets the larger 
 dispersion in the PL relation in $\wi$ for the SMC compared to the
LMC in terms of a larger intrinsic depth of the former Cloud.

For both the SMC and the LMC,
the predicted $\log \, P$ - $W_{\rm I}$ mean relationships are shifted
compared to the observed relationships. Since the shift between
observed and predicted relationships 
is {\it almost constant} on the whole range of $P$
 and is the {\it same} between F and 1H relations, we find that
an increase
of the distance moduli $\mu_0$ 
by the same amount $\sim$ 0.15 - 0.2 mag for both clouds 
removes easily this discrepancy. This is illustrated in
the insets of Figs. \ref{fig4} and \ref{fig5}.
Note that such an increase of
$\mu_0$ does not alter the agreement found between observations
and models in the $I$-band (see Fig. 2 and Fig. 3).
It however increases the discrepancy in the $V$-band
 mentioned in the previous
section. 
It is difficult to analyse the reason of such inconsistency. 
The choice of the reddening and of the extinction curve
coefficients, affecting mostly the $V$-band,
could be the reason of this problem. This  emphasizes
the large uncertainties intrinsic to observations in this filter
and suggests to use rather $W_{\rm I}$ to derive a distance modulus.
Note however that, independently of reddening and distance modulus,
the models do predict the correct
 location of the observed  1H sequence
{\it relative} to the F sequence, for both SMC and LMC.

We also note that the models yield a difference of distance moduli
between SMC and LMC of $\sim 0.45$, in good agreement with 
previous determinations based on Cepheids (Laney and Stobie 1994; Udalski et al. 
1999c), RR Lyr or Red Clump stars (cf. Udalski et al. 1999c; Udalski
2000). Finally, Caputo et al. (2000) also derived theoretical 
$\log \, P$ - $W_{\rm I}$  relationships for the fundamental mode which
are shallower than the present relationships. Alibert et al. (1999)
already mentioned and discussed  this difference 
for fundamental mode period - magnitude relationships.
As illustrated in Figs. \ref{fig4} and \ref{fig5},
the Caputo et al. (2000) $\log \, P$ - $W_{\rm I}$ relationships are also shallower 
compared to the observed
data in both SMC and LMC. For the distance moduli
 adopted in the insets of Figs. \ref{fig4} and \ref{fig5}, the F sequence of
Caputo et al. (2000) reproduce the F data at long
period, but merges into the observed 1H pulsators at shorter periods. 
A variation of the distance modulus cannot clearly solve
such discrepancy. This highlights the constraint on models and
distance moduli provided by the combination of F and 1H observed sequences.  


\section{Uncertainties on distance determination}

Alibert et al. (1999) analysed the effects of 
uncertainties inherent to stellar evolution models: convection
treatment, overshooting, mass loss, initial helium abundance,  etc...
The main conclusion is that 
such uncertainties barely affect the PL relationships and the conclusion 
on metallicity effects. We however recall that such uncertainties can
affect the comparison between models and observations
if the {\it mass} is used as another constraint. 

The main uncertainty
in our calculations is due to the neglect of convection -
pulsation coupling in our linear stability analysis. Convection is frozen
in, which means that the perturbation $\delta F_{\rm conv}$ of the
convective
flux is neglected in the linearized energy equation.
An arbitrary criterion is used to define a red edge for the
instability strip (see Alibert et al. 1999 for details), since
 such approximation
 cannot yield naturally a red edge.


A time dependent non local theory of convection is required to take into
account the effect of convection on pulsation. 
Such a theory is lacking and the 
current recipes include several free parameters (cf. Yecko et al. 1998 and
references therein). 
Despite the lack of a robust theory, it has now become clear that convective
energy transport is essential to describe the observed properties
of pulsating stars such as light curves or Fourier coefficients
 (Yecko et al. 1998; Feuchtinger 1999; Feuchtinger
et al. 2000). 

In a systematic analysis based on linear stability
analysis of Cepheid models,  
Yecko et al. (1998) have  shown the
high sensitivity of the position of the IS to these free parameters, which need
to be calibrated according to astronomical observations. But as emphasized
by Feuchtinger et al. (2000) even a correct description of the instability
strip requires hydrodynamical calculations, since the red edge
is determined by nonlinear effects. Even for  the blue edge,
Feuchtinger et al. (2000) report a shift toward higher
$\te$ by about 350 K for the fundamental mode when convection is
included in the pulsation calculation.  

In the present paper, we derive a rough  estimate of  
the sensitivity of the period - magnitude relationships derived
under the afore-mentioned approximations, and the effects on distance
determination. A more detailed analysis and 
the quantification of uncertainties resulting from
convection is addressed in a forthcoming paper (Alibert \& Baraffe
2001).
Since our results yield a general good agreement with the
width of observed instability strips (see Alibert et al. 1999),
a first estimate of such uncertainties can easily be derived by
shifting
arbitrarily both the blue and red edges of the instability strip
toward cooler or hotter $\te$, keeping the width
unmodified. In the following, we focus on fundamental PL
relationships,
since they are mostly used for distance determinations.
Inspired by the afore-mentioned results by Feuchtinger et al. (2000),
we adopt a shift in $\te$ of 350 K.
Based on the models appropriate for the LMC ($Z$ = 0.01),
a shift by 350 K toward hotter (cooler)
 $\te$ yields brighter (fainter) magnitudes
at a given $P$.  The effect decreases toward near-IR filters: 
the resulting variation of the magnitude in the $B$-band reaches up to 0.6 mag,
it does not exceed 0.3 mag in the $I$-band and remains below 0.2 mag at
longer wavelengths.
This is expected,
given the decreasing sensitivity to $\te$ of the flux toward
longer wavelengths. In the $V$-band,
$\mv$ varies from 0.3 mag at short $P$ ($\log P = 0.5$) up to
0.45 mag at longer $P$ ($\log P = 1.7$). In the $K$ band, the
effect is less than 0.1 mag on the whole range of periods. 
This simple test highlights the high sensitivity
of distance determinations based on theoretical 
$P$ - $\mv$ relationships, since a variation of 350 K on the
location of the blue or red edge can easily
result from uncertainties due to convection (see Yecko et al. 1999; 
Feuchtinger et al. 2000; Alibert
and Baraffe 2001). 

\begin{figure}
\psfig{file=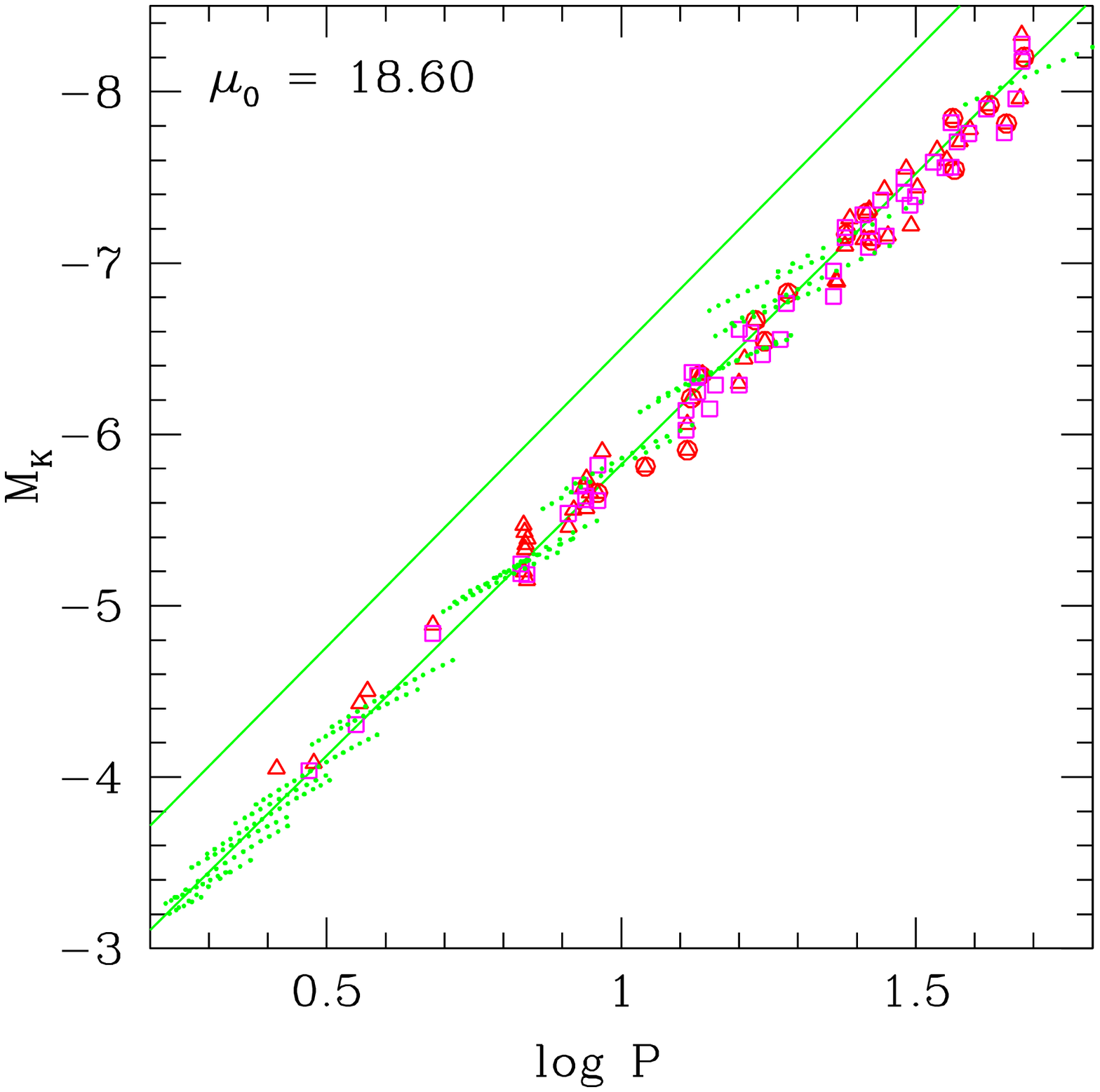,height=110mm,width=88mm}
\caption{$\log P$ (in days)  - $\mk$ diagram  for 
LMC observed Cepheids.
The dots correspond to models for fundamental
mode pulsators during core He burning,  with $Z$ = 0.01.
The open squares are from Persson et  al. (2000).
The triangles are observations from Gieren et al. (1998).
The open circles are from Laney and Stobie (1994).
The 2 solid curves are the theoretical mean PL relationships for respectively
F (lower curve) and 1H (upper curve).
}
\label{figlmc}
\end{figure}

Finally, since our simple test clearly shows that near-IR period - magnitude
 relationships are more reliable, it is worth to 
derive a distance modulus for LMC based on near-IR data.
Recent observations in the near-IR 
(Madore, 2000 priv. comm.; Persson et al. 2000) 
complete data already available from Laney and Stobie (1994) and
Gieren et al. (1998).  
The $P$ - $\mk$ relationship (Table 2)
yields a distance modulus for LMC of 18.60- 18.70, as illustrated in
Fig. \ref{figlmc}.
Interestingly enough, 
this distance modulus is in good agreement with
the value determined in \S 3.3 based on $\wi$ and the combination of
F and 1H  observed sequences.

\section{Conclusion}

We have derived period - magnitude relationships for 1H pulsators
in different filters. In contrast to fundamental
mode, we find a small effect of metallicity
in PL relationships for first overtone pulsators. This effect is
due to the dependence on metallicity of the ratio
$P_1$/$P_0$. 
Our models reproduce the location of
both F and 1H observed sequences of OGLE2 in the SMC and
the LMC. Using the reddening free index $\wi$, models
and observations are in good agreement for a LMC distance
modulus $\mu_0$ = 18.65 - 18.70. We note however
an inconsistency since this value yields a significant
discrepancy between predictions and data in the $V$-band.
Such inconsistency may illustrate problems in the choice
of the extinction coefficients and the reddening, which
are important for the comparison between models and data in the $V$-band
but not crucial for $\wi$.

We show that an arbitrary shift of 350 K in $\te$ of the location of the
instability strip yields up to 0.45 mag effect on $\mv$ at a given
$P$. Toward near-infrared wavelengths, the effect is smaller
and less than 0.1 mag in the $K$-band. 
Using recent Cepheid
data in {\it the $K$-band for fundamental pulsators}  in the LMC,
the models predict
a distance modulus for LMC $\mu_0$ = 18.60 - 18.70, in good agreement
with the predictions based on {\it $W_I$ and combined 1H and F data}.


Although Cepheid observations in the $V$-band are widely used 
for distance determination, we stress
 the high uncertainties inherent to observations in this filter
(reddening correction, intrinsic dispersion in the instability strip,
etc...) and the high sensitivity of
 $P$ - $\mv$ relationships to theoretical uncertainties. 
We also emphasize that even if the coupling between convection
and pulsation provides a large source of uncertainty
in the present work and in all current theoretical calculations,
predictions in the $K$-band are expected to be much
less affected by such uncertainty.


\begin{acknowledgements} We are indebted to Andrzej Udalski for providing
data and results prior to publication. 
We are grateful to Barry Madore for valuable discussions and
for making available to us near-IR data for the LMC.
Many thanks to our referee, A. Gautschy, who contributed
to the improvement of the manuscript and provided excellent ideas.
The calculations were performed using facilities at Centre
d'Etudes Nucl\'eaires de Grenoble.
\end{acknowledgements}


\begin{thebibliography}{}

\bibitem[]{} Afonso, C. et al. (EROS) 2000, \aaps, submitted,
astro-ph/9907355

\bibitem[]{} Alcock, C. et al. (MACHO) 1997, ApJ, 486, 697

\bibitem[]{} Alibert, Y., Baraffe, I., Hauschildt, P.H., Allard, F. 1999, \aap,
344, 551

\bibitem[]{} Alibert, Y., Baraffe, I. 2001, in preparation

\bibitem[1998]{baraffe98} Baraffe, I., Alibert, Y., M\'era, D., Chabrier, G.,
Beaulieu, J-P., 1998, \apj, 499, L205

\bibitem[1999]{bauer} Bauer, F., et al. 1998, \aap, 348, 175 

\bibitem[1990]{bessell} Bessell, M.S., 1990, PASP, 102, 1181

\bibitem[]{} Caputo, F., Marconi, M., Musella, I. 2000, \aap, 354, 610



\bibitem[]{} Cox, J.P. 1980, {\it Theory of Stellar Pulsations}, Princeton University Press, p. 108

\bibitem[]{} Epstein, I. 1950, \apj, 112, 6

\bibitem[]{} Feuchtinger, M.U.  1999, \aaps, 136, 217

\bibitem[]{} Feuchtinger, M.U., Buchler, J.R., Kollath, Z. 2000, \apj,
accepted, astro-ph/0005230

\bibitem[]{} Freedman, W. 2000, `David Schramm Memorial Volume of
Physics Reports', astro-ph/9909076

\bibitem[1998]{gieren98} Gieren, W.P., Fouqu\'e, P., Gomez, M., 1998,
\apj, 496, 17

\bibitem[]{} Groenewegen, M.A.T., 2000, \aap, accepted, astro-ph/0010298

\bibitem[1994]{laney} Laney, C.D.,  Stobie, R.S. 1994, \mnras, 266, 441 (LS94)
\bibitem[1992]{leggett} Leggett, S.K., 1992, \apjs, 82, 351

\bibitem[]{} Madore, B. F., Freedman, W.L. 1991, PASP, 103, 933

\bibitem[1996]{renault} Renault, C., et al. 1996, 12th IAP Astr. meeting,
``Astrophysical returns of microlensing surveys'',
eds R. Ferlet and JP. Maillard

\bibitem[]{} Persson, E. et al. 2000, in preparation
 

\bibitem[]{} Schlegel, D.J., Finkbeiner, D.P., Davies, M. 1998, \aj, 500, 525

\bibitem[]{} Udalski, A. 2000, AcA, 50, 279

\bibitem[1997]{udalski} Udalski, A., Kubiak, M., Szymanski, M., 1997,
AcA, 47, 319


\bibitem[]{} Udalski, A., Soszynski, I., Szymanski, M., Kubiak, M., Pietrzynski, G., 
Wozniak, P., Zebrun, K. 1999a, AcA, 49, 223

\bibitem[]{} Udalski, A., Soszynski, I., Szymanski, M., Kubiak, M., Pietrzynski, G., 
Wozniak, P., Zebrun, K. 1999b, AcA, 49, 437

\bibitem[]{} Udalski, A. Szymanski, M., Kubiak, M., Pietrzynski, G., Soszynski, I.,
Wozniak, P., Zebrun, K. 1999c, AcA, 49, 201

\bibitem[]{} Whitney, C.A. 1983, \apj, 274, 830

\bibitem[1998]{yecko} Yecko, P.A., Kollath, Z., Buchler, J.R., 1998, \aap,
336, 553

\end{thebibliography}
\end{document}